\documentclass[sigconf]{acmart}

\usepackage{amsmath,amssymb,amsfonts}
\usepackage{algorithmic}
\usepackage{graphicx}
\usepackage{textcomp}
\usepackage{booktabs}
\usepackage{multirow}
\usepackage{bm}
\usepackage[justification=centering]{caption}
\usepackage{balance}
\usepackage{float}

\AtBeginDocument{%
  \providecommand\BibTeX{{%
    \normalfont B\kern-0.5em{\scshape i\kern-0.25em b}\kern-0.8em\TeX}}}

\copyrightyear{2021}
\acmYear{2021}
\setcopyright{acmcopyright}\acmConference[EASE 2021]{Evaluation and Assessment
in Software Engineering}{June 21--23, 2021}{Trondheim, Norway}
\acmBooktitle{Evaluation and Assessment in Software Engineering (EASE 2021), June
21--23, 2021, Trondheim, Norway}
\acmPrice{15.00}
\acmDOI{10.1145/3463274.3463325}
\acmISBN{978-1-4503-9053-8/21/06}

\begin{document}

\title{A Machine Learning Based Ensemble Method for Automatic Multiclass Classification of Decisions}



\author{Liming Fu, Peng Liang$^{*}$, Xueying Li}
\affiliation{%
  \institution{School of Computer Science, Wuhan University}
  \institution{Wuhan, China}
  \institution{\{limingfu, liangp, xueyingli\}@whu.edu.cn}
  \country{}
}

\author{Chen Yang}
\affiliation{%
  \institution{IBO Technology (Shenzhen) Co., Ltd.}
  \institution{Shenzhen, China}
  \institution{cyang@whu.edu.cn}
  \country{}
}

\renewcommand{\shortauthors}{L. Fu et al.}

\begin{abstract}
Stakeholders make various types of decisions with respect to requirements, design, management, and so on during the software development life cycle. Nevertheless, these decisions are typically not well documented and classified due to limited human resources, time, and budget. To this end, automatic approaches provide a promising way. In this paper, we aimed at automatically classifying decisions into five types to help stakeholders better document and understand decisions. First, we collected a dataset from the Hibernate developer mailing list. We then experimented and evaluated 270 configurations regarding feature selection, feature extraction techniques, and machine learning classifiers to seek the best configuration for classifying decisions. Especially, we applied an ensemble learning method and constructed ensemble classifiers to compare the performance between ensemble classifiers and base classifiers. Our experiment results show that (1) feature selection can decently improve the classification results; (2) ensemble classifiers can outperform base classifiers provided that ensemble classifiers are well constructed; (3) BoW + 50\% features selected by feature selection with an ensemble classifier that combines Naïve Bayes (NB), Logistic Regression (LR), and Support Vector Machine (SVM) achieves the best classification result (with a weighted precision of 0.750, a weighted recall of 0.739, and a weighted F1-score of 0.727) among all the configurations. Our work can benefit various types of stakeholders in software development through providing an automatic approach for effectively classifying decisions into specific types that are relevant to their interests.
\end{abstract}


\maketitle

\section{Introduction} \label{Introduction}
Stakeholders make numerous decisions with respect to software design, requirements, management, and so on during the software development life cycle. These decisions potentially affect the functional and non-functional properties of software systems \cite{shahbazian2018recovering} and thus have a crucial impact on software development and maintenance. Considering the importance of decisions, it is essential for stakeholders to document and understand the past decisions. Documenting these decisions will help stakeholders organize development knowledge and reduce its vaporization, therefore controlling the development process and maintenance costs \cite{van2006design}. In addition, it also contributes to rationale management through recording decisions. The availability of rationale not only increases the developers' understanding of the system, but also facilitates the training of new members \cite{dutoit2006rationale}.


However, due to limited human resources, time, and budget, decisions are rarely well documented in Open Source Software (OSS) \cite{ding2014open} and industrial projects \cite{ambler2002agile}. The loss of this information can thus hurt the development and maintenance of software systems \cite{shahbazian2018recovering}. Even though decisions are not explicitly documented, they may be implicitly captured in many tools, such as mailing lists, issue tracking systems, and version control systems, which are used for stakeholders to communicate with each other about the issues encountered in software development \cite{liautomatic}. Stakeholders can learn diverse decision knowledge accumulated during the process of project development, such as requirements, design, and testing, from these discussions \cite{li2019decisions}. It is thus desirable to identify and classify decisions implicitly captured in these tools for documentation purposes. However, there are massive textual artifacts in these tools, which may also contain lots of meaningless information. Manually identifying and classifying decisions is rather time-consuming, and therefore it may not be applicable in practice. To this end, exploiting automatic methods to identify and classify decisions in software development is a promising way. 


While several studies have been conducted to use automatic methods to mine decisions in software development, most of them focus on either identifying or classifying specific type of decisions (e.g., design decisions in \cite{bhat2017automatic}) or identifying decisions without further classification (e.g., \cite{liautomatic}). To the best of our knowledge, limited work has been done on classifying decisions into types that are beneficial for various types of stakeholders in development. Stakeholders are engaged with different tasks in software development, thus they are more concerned about the decisions related to their interests. For example, developers may highlight decisions concerning code issues. Considering this, we planned to automatically classify decisions into five decision types (i.e., Design, Requirement, Management, Construction, and Testing Decision) according to the types got in our recent work \cite{li2019decisions}.

In this work, we first collected our dataset from the Hibernate developer mailing list. We then examined the performance of a feature selection algorithm (i.e., CHI$^2$), three feature extraction techniques, and four base machine learning (ML) classifiers. Moreover, we used an ensemble method and constructed five ensemble classifiers using the aforementioned four base ML classifiers. Ensemble methods that train multiple classifiers and then combine them are a kind of state-of-the-art learning approach. It is well known that an ensemble classifier is usually significantly more accurate than a single classifier \cite{zhou2012ensemble}. Overall, 270 configurations regarding ten different percentages of selected features in feature selection, three feature extraction techniques, and nine classifiers were evaluated through weighted precision, recall, and F1-score. The contribution of this work is threefold: (1) the first empirical study focusing on automatic classification of decisions into five decision types that are beneficial for various types of stakeholders in development; (2) an effective automatic approach with ensemble learning to classify decisions into five decision types with a weighted precision of 0.750, a weighted recall of 0.739, and a weighted F1-score of 0.727; (3) empirical evidence about the effectiveness comparison between using base supervised machine learning classifiers and ensemble classifiers to automatically classify decisions in software development.

The rest of this paper is structured as follows: Section \ref{Related Work} introduces the related work on decisions in software development and automatically mining textual artifacts in software development. Section \ref{Research Design} describes the research design and Section \ref{Experiment} presents the detailed procedures of the experiment. The results and analysis of our experiment are provided in Section \ref{Results and Analysis}. Section \ref{Discussion} discusses the feature terms and the implications of our study. Section \ref{Threats to Validity} clarifies the threats to the validity. Finally, we conclude this work with future research directions in Section \ref{Conclusions and Future Work}.

\section{Related Work} \label{Related Work}

\subsection{Decisions in Software Development}\label{Decisions in Software Development}
Decisions have a crucial effect on development and maintenance of software systems. Many researches have focused on analyzing decisions and their making in software development. For instance, Razavian \textit{et al.} systematically analyzed 38 research papers on architecture decision-making with a special focus on human aspects of architecture decision-making \cite{razavian2019empirical}. Their study highlights that decisions as well as the decision-making process are important knowledge in architecture design. Mendes \textit{et al.} aimed at studying how the personality of decision makers can influence their decision-making \cite{mendes2019relationship}. They reported 75 relationships between 28 different personality aspects and 30 different decision-making aspects, which can help in improving how a decision is made in software engineering context.

There are also a few researches focusing on employing automatic approaches in decision-related work. Bhat \textit{et al.} proposed a two-phase supervised machine learning based approach to automatically detect design decisions from issues in issue tracking systems of two large OSS projects, and then classify the detected design decisions into three categories (i.e., Behavioral decision, Structural decision, and Ban decision) \cite{bhat2017automatic}. In their follow-up work, they developed a tool named ADeX to automatically extract, enrich, and generate specific views on architectural knowledge to support architects’ decisions-making process \cite{bhat2019adex}. Shahbazian \textit{et al.} developed a technique called RecovAr to automatically recover design decisions from the readily available history artifacts of projects (e.g., an issue tracker and version control repository) \cite{shahbazian2018recovering}. Their work provides a way that developers can follow to preserve design decision knowledge in their projects. Li \textit{et al.} conducted an experiment with 160 configurations to automatically identify decisions from an OSS mailing list \cite{liautomatic}. The evaluation results show that their approach can effectively identify decisions from the mailing list.

Inspired by the work above (i.e., \cite{bhat2017automatic,bhat2019adex,shahbazian2018recovering,liautomatic}), we are motivated to develop an automatic approach to help researchers and practitioners better understand specific types of decisions in software development. The difference between our work and existing work is that we considered classifying decisions into five decision types that are beneficial to various types of stakeholders in software development.

\subsection{Automatically Mining Textual Artifacts in Software Development}\label{Automatic Approaches for Mining Valid Information from Textual Artifacts}
Automatically mining textual artifacts is subject to widespread attention in software development. Besides the studies that focus on decisions, many studies have also focused on automatically mining artifacts that are highly related to decisions. Li \textit{et al.} used Word2Vec to extract features and trained seven machine learning classifiers for automatically identifying assumptions from an OSS developer mailing list \cite{li2019automatic}. Their experiment results show that the automatic approach can effectively identify assumptions by using the SVM algorithm with a precision of 0.829, a recall of 0.812, and an F1-score of 0.819. Rogers \textit{et al.} compared text mining and text parsing techniques to automatically mine rationale from bug reports \cite{rogers2015using}, and they argued that the results obtained by their approach are comparable in accuracy to those obtained by human annotators.

Automatic approaches have also been applied to mine other textual artifacts. Shah \textit{et al.} considered automatically classifying user reviews into feature evaluation, feature requests, and bug reports \cite{shah2018simple}. Their results show that the simple BoW model is competitive compared with complex models adopting rich linguistic features. Huang \textit{et al.} proposed an automated approach to detect self-admitted technical debt from source code comments of eight OSS projects \cite{huang2018identifying}. Their approach trained an individual sub-classifier for each source project, and then combined them as an ensemble classifier to vote for the comments in the target project. The results indicate that the ensemble classifier could improve F1-score. 

The above automation related work (both on decisions and other textual artifacts) motivates us to use an automatic approach to classify decisions from OSS developer mailing lists. Their experiment setups lay the foundation for our experiment.

\section{Research Design} \label{Research Design}
In this section, we first formulate the goal with three Research Questions (RQs) that we try to answer through this study. We then briefly describe the five decision types. Finally, we introduce feature selection, feature extraction techniques, machine learning classifiers, and ensemble classifiers that are utilized for classifying decisions in the experiment.

\subsection{Goal and Research Questions}\label{Goal and Research Questions}
The goal of this work is to analyze textual information in the Hibernate developer mailing list \textbf{for the purpose of} automatic classification \textbf{with respect to} various types of decisions in software development \textbf{from the point of} view of OSS developers \textbf{in the context of} OSS development, which is formulated through the Goal Question Metric approach \cite{caldiera1994goal}. This work comprises three RQs, which are presented in the following.

\textbf{RQ1. Does feature selection improve the classification results?}

\textbf{RQ2. Do the ensemble classifiers outperform the base classifiers for classifying decisions?}

\textbf{RQ3. What is the best configuration for classifying decisions into five decision types?}

\subsection{Types of Decisions}\label{Types of Decisions}
In our previous study on decisions and their making in software development \cite{li2019decisions}, we analyzed 980 decisions from 9006 posts of the Hibernate developer mailing list and identified five decision types (i.e., Design, Requirement, Management, Construction, and Testing Decision) from those decisions using grounded theory. We adapted and adopted the description of the five decision types in this work based on our previous description in \cite{li2019decisions}:

\textbf{Design Decisions} are decisions made about the design of software. Design decisions are related to various elements in software systems, such as components, interfaces, layers, and subsystems. They indicate the specification of the composition, properties, structures, behaviors, or relationships of these elements, or reflect the rules, patterns, or constraints that these elements should follow.

\textbf{Requirement Decisions} are related to functional requirements or non-functional requirements of software systems. They aim to provide or improve functionality for software, or meet a set of quality attributes that the software should exhibit, such as reliability, portability, and performance.

\textbf{Management Decisions} are related to project management of software. They indicate the maintenance and improvement in documentations of software, version control of software, development plans of software, and other management operations.

\textbf{Construction Decisions} are related to the specific implementation of software. They focus on the source code of project such as source code annotation, the configuration files of the project, and other specific operations for the project.  

\textbf{Testing Decisions} are related to the test and debugging of software. They cover the test behavior, the creation or change of test cases, or the bug report of software.

We found that the above five decision types can basically cover the decisions made during the development and maintenance of Hibernate \cite{li2019decisions}. Therefore, we considered classifying decisions from the Hibernate developer mailing list into these five decision types. The decision examples of these types are provided in Table \ref{decision-examples}.  

\begin{table*}[htb]
\centering
\caption{Examples of Each Decision Type from the Hibernate Developer Mailing List}
\label{decision-examples}
\begin{tabular}{@{}llll@{}}
\toprule
\textbf{Type}         &  & \textbf{Examples}                                                                             &  \\ \midrule
Design Decision &  & ``\textit{I have redefined what happens when a persistent object holds a reference to a transient object at flush() time.}'' &  \\ \midrule
Requirement Decision  &  & ``\textit{I am adding support for template-based functions as part of the EJBQL issue.}''     &  \\ \midrule
Management Decision   &  & ``\textit{We will release Hibernate 3.0alpha early next week.}                                &  \\ \midrule
Construction Decision &  & ``\textit{I added a new instance array to store the Types of each of the columns.}''          &  \\ \midrule
Testing Decision      &  & ``\textit{I will run the tests on all platforms before the next release just to make sure.}'' &  \\ \bottomrule
\end{tabular}
\end{table*}

\subsection{Feature Selection} \label{Feature Selection}
Terms that exist in more types contain less type information, and consequently they are less important for classification \cite{lu2017automatic}. Moreover, research shows that feature selection can improve the performance of text classification \cite{yang1997comparative}. Therefore, we apply feature selection to identify a subset of features (i.e., terms) that provide expected discrimination between the types of decisions and thus are useful for classification.

There are several frequently used feature selection algorithms, such as Mutual Information, Information Gain, and Chi Squared (i.e., CHI$^2$) \cite{forman2003extensive}. Among them, CHI$^2$ is one of the most efficient methods used in many studies related to text classification tasks \cite{guzman2015ensemble,lu2017automatic,rustam2020classification}. The CHI$^2$ value of term \bm{$T_i$} for a certain decision type (\bm{$Type_k$}) is calculated through Formula (\ref{chi2-formula}), where \bm{$N$} denotes the total number of decisions in our dataset, variable \bm{$a$} denotes the number of decisions that have term \bm{$T_i$} and belong to \bm{$Type_k$}, variable \bm{$b$} denotes the number of decisions that have term \bm{$T_i$} and do not belong to \bm{$Type_k$}, variable \bm{$c$} denotes the number of decisions that do not have term \bm{$T_i$} and belong to \bm{$Type_k$}, and variable \bm{$d$} denotes the number of decisions that do not have term \bm{$T_i$} and do not belong to \bm{$Type_k$}. The greater the value of $CHI^{2}(T_i,Type_k)$ is, the more type information the Term $T_i$ contains.

\begin{equation}
\label{chi2-formula}
	CHI^{2}(T_i,Type_k) = \frac{N*(ad-bc)^2}{(a+c)(b+d)(a+b)(c+d)}
\end{equation}

In this work, we calculated the CHI$^2$ value for each term in each decision type (after decisions preprocessing). Then for each decision type, the terms were ranked by the CHI$^2$ value in a descending order and we chose the top $n$ percent of the terms as feature terms. The value of $n$ is a threshold that depends on the specific classification task and the dataset, and it ranges from 1\% to 100\%. The selection result of value of \bm{$n$} is described in Section \ref{experiment fs}. 


\subsection{Feature Extraction Techniques}\label{Feature Extraction Techniques}
In text classification tasks, texts are modeled as vectors that can represent their features. Three well-performed techniques are widely used in text classification tasks for extracting text features. Therefore, we also experimented and compared these three techniques in this work, which are briefly described below.

\textbf{BoW:} Bag of Words, builds a vocabulary composed of all unique terms in the decisions of our dataset and then represents each decision by counting the term frequency of every term in the vocabulary. 

\textbf{TF-IDF:} Term Frequency - Inverse Document Frequency, uses the same textual features (i.e., terms) as BoW does. But different from BoW, TF-IDF evaluates a term's weight by combining term frequency with the inverse document frequency, which implies that the importance of a term increases with the term frequency in a sentence, but decreases inversely with the frequency of its appearance in the whole corpus.

\textbf{Word2Vec:} is prevalently used in the Nature Language Processing (NLP) community. It can learn semantic knowledge from a large number of text corpus and then transform words in the corpus into vectors with rich semantic information \cite{rong2014word2vec}. Continuous Bag-of-Words (CBoW) and Continuous Skip-Gram (Skip-Gram) are two models of Word2Vec, which are both efficient methods for various NLP tasks, such as text classification. CBoW can predict the current word through the surrounding context while Skip-Gram can forecast the context according to the current words. In this work, we use CBoW model to generate word embeddings and represent each decision with a vector calculated through Formula (\ref{word2vec-formula}), where \bm{$d_i$} denotes the decision \bm{$i$}, \bm{$V_{d_{i}}$} denotes the vector representation of \bm{$d_i$}, \bm{$N$} denotes the total number of words in \bm{$d_i$}, \bm{$d(i,n)$} denotes the \bm{$n^{th}$} word in \bm{$d_i$}, and \bm{$W_d(i,n)$} denotes the word embeddings of \bm{$d(i,n)$} generated by Word2Vec.

\begin{equation}
\label{word2vec-formula}
	V_{d_{i}}=\frac{1}{N}\sum_{n=1}^N W_{d(i,n)}
\end{equation}

\subsection{Machine Learning Classifiers} \label{Machine Learning Classifiers}
A recent research shows that some supervised ML classifiers, such as SVM or Random Forest (RF), can perform equally well or even better than deep learning techniques in text classification tasks \cite{fu2017easy}. Also, deep learning techniques usually are more complex, slower, and tend to over-fit the models when a small dataset is used \cite{beyer2020kind}. Considering this, we decided to utilize supervised ML classifiers in our experiment. Many studies show that different ML classifiers can lead to various performance in different classification tasks and each ML classifier may have its own suitable software artifacts for achieving the best classification results. For instance, NB has been reported to outperform other machine learning algorithms in the automatic classification of app reviews \cite{maalej2016automatic}. LR achieves the best performance when identifying code-smell discussions from SO \cite{shcherban2020automatic}. SVM gets the best classification result when identifying assumptions and decisions from the Hibernate developer mailing list \cite{li2019automatic,liautomatic}. Therefore, four widely used machine learning classifiers (i.e., NB, LR, SVM, and RF) are used as base classifiers in our experiment to seek the best configuration to classify decisions.

\subsection{Ensemble Classifiers} \label{Ensemble Classifier}
Ensemble learning is getting more and more attention in recent years and it has been widely used in many areas to solve classification problems. In contrast to base machine learning approaches, which construct one classifier from the training data, an ensemble classifier constructs a set of base classifiers and combines predictions from base classifiers into the final prediction to solve the same problem \cite{zhou2012ensemble}. Dietterich considered that it is often possible to create excellent ensemble classifiers and attributed the benefits of ensemble classifiers to three fundamental reasons: Statistical, Computational, and Representational \cite{dietterich2000ensemble}. Besides, many research results also show that ensemble classifiers can obtain better performance than base classifiers \cite{huang2018identifying,guzman2015ensemble,da2014tweet}, which motivates us to apply ensemble methods in our experiment and compare the performances between base classifiers and ensemble classifiers when classifying decisions. 

However, effective ensemble classifiers require that the base classifiers can exhibit some level of diversity \cite{da2014tweet}. Brown \textit{et al.} suggested a taxonomy of methods for creating diversity in ensemble classifiers: (1) varying the start points in the hypothesis space; (2) varying the set of accessible hypothesis (e.g., by manipulating the training set for base classifiers); and (3) varying the way each base classifier traverses the space of possible hypotheses (e.g., by using different base classifiers or combination strategies) \cite{brown2005diversity}. In this work, we focus on the third method and use different combinations of four base classifiers (see Section \ref{Machine Learning Classifiers}) to construct five ensemble classifiers for classifying decisions. Also, soft voting, one of the most popular and fundamental combination methods \cite{zhou2012ensemble}, is chosen as the combination method for the five ensemble classifiers. It averages probabilities obtained by base classifiers for each class, and then outputs the class with the maximum average probability as the final prediction result. Table \ref{ens-classifiers} presents the detailed information of these ensemble classifiers, in which SVE denotes Soft Vote Ensemble.

Note that there is no guarantee that ensemble classifiers will always perform better than base classifiers. Therefore, we used the five ensemble classifiers to seek whether the ensemble classifiers can outperform the base classifiers when classifying decisions.

\begin{table}[h]
\caption{Combinations of Base Classifiers for Ensemble Classifiers}
\label{ens-classifiers}
\centering
\begin{tabular}{@{}llll@{}}
\toprule
 & \textbf{Ensemble classifiers} & \textbf{Combination of base classifiers} &  \\ \midrule
 & SVE1 & NB, LR, SVM, and RF  &  \\
 & SVE2 & NB, LR, and SVM      &  \\
 & SVE3 & LR, SVM, and RF      &  \\
 & SVE4 & NB, SVM, and RF      &  \\
 & SVE5 & NB, LR, and RF       &  \\ \bottomrule
\end{tabular}
\vspace{-1.5em}
\end{table}

\section{Experiment}\label{Experiment}
In this section, we present in detail how we designed our experiment to automatically classify the decisions from the Hibernate developer mailing list. We first specify the process of the decision classification experiment, and then describe each phase in the process. Finally, we provide the methodology to evaluate the performance of classification results.

\subsection{Decision Classification Process}\label{Decision Classification Process}
In this section, we show the execution process we followed and the techniques we employed to conduct our experiment. Specially, the process consists of six phases:

\textbf{Phase 1: Collect Decisions:} We collected decisions from the Hibernate developer mailing list.

\textbf{Phase 2: Preprocess Decisions:} We preprocessed the collected decisions by eliminating useless textual characters, carrying out lemmatization, and removing stop words.

\textbf{Phase 3: Select Features:} We applied CHI$^2$ feature selection algorithm to identify a subset of features that are most useful in differentiating decision types. 

\textbf{Phase 4: Extract Features:} We applied three feature extraction techniques (i.e., BoW, TF-IDF, and Word2Vec) to extract textual features and calculate the vector value of each decision.

\textbf{Phase 5: Train Classifiers:} We employed four machine learning classification algorithms (i.e., NB, LR, SVM, and RF) to separately train base classifiers and then we trained five ensemble classifiers constructed by the aforementioned four base classifiers. After that, decisions would be classified by the four trained base classifiers and the five trained ensemble classifiers.

\textbf{Phase 6: Evaluate Trained Classifiers:} We used weighted precision, weighted recall, and weighted F1-score to evaluate the performance of the trained classifiers in \textbf{Phase 5}.

\subsection{Decisions Collection}\label{Decisions Collection}
In our recent study, we conducted a replication experiment for automatically identifying decisions from the Hibernate developer mailing list \cite{fuliautodecision}. The dataset used in the replication experiment contains 844 decisions. In this work, we reused the 844 decisions and manually labelled these decisions into five decision types. To mitigate any personal bias in creating the dataset, we first conducted a pilot labelling with 300 decisions by the first and third authors independently. Conflicts were discussed and resolved with the second author to make sure that they had a consistent understanding of the criteria of data labelling. The first and third authors then labelled the remaining decisions independently. Disagreements on labelled decisions were still discussed and resolved with the second author. We measured the inter-rater reliability and calculated Cohen’s Kappa coefficient \cite{cohen1960coefficient} as a way to verify the consistency between the labelled decisions by the first and third authors. The result is 0.856, which indicates that the two authors reached a good agreement. The detailed statistics of our dataset is presented in Table \ref{tab-decision dataset}. In the meanwhile, the dataset has been provided online \cite{replication-package}.

\begin{table}[htb]
\centering
\caption{Numbers and Proportions of Manually Labelled Decisions in the Dataset}
\label{tab-decision dataset}
\begin{tabular}{@{}lllllllllll@{}}
\toprule
\multicolumn{3}{l}{\textbf{Type}}                  &  & \multicolumn{3}{l}{\textbf{Number}} &  & \multicolumn{3}{l}{\textbf{Proportion}} \\ \midrule
\multicolumn{3}{l}{Design Decision}       &  & \multicolumn{3}{l}{314}        &  & \multicolumn{3}{l}{0.3720}      \\
\multicolumn{3}{l}{Requirement Decision}  &  & \multicolumn{3}{l}{299}        &  & \multicolumn{3}{l}{0.3543}      \\
\multicolumn{3}{l}{Management Decision}   &  & \multicolumn{3}{l}{109}        &  & \multicolumn{3}{l}{0.1291}      \\
\multicolumn{3}{l}{Construction Decision} &  & \multicolumn{3}{l}{99}        &  & 
\multicolumn{3}{l}{0.1173}      \\
\multicolumn{3}{l}{Testing Decision}         &  & \multicolumn{3}{l}{23}         &  & \multicolumn{3}{l}{0.0273}      \\
\multicolumn{3}{l}{Total}                 &  & \multicolumn{3}{l}{844}        &  & \multicolumn{3}{l}{1.0000}      \\ \bottomrule
\end{tabular}
\vspace{-1.5em}
\end{table}

\subsection{Decisions Preprocessing}\label{Decisions Preprocessing}
The process of text preprocessing is composed of three steps:

\textbf{1) Eliminating Useless Textual Characters:} Due to the fact that Hibernate developers may use natural language to communicate with other developers through mails, there may be a lot of useless characters in the decision sentences, such as ``:)'' and ``*''. We only need to consider terms that can provide valid information for classifying decisions and thus may improve classification performance. Therefore, we eliminated useless characters and numbers, and only retained valid terms.

\textbf{2) Lemmatization:} Lemmatization is one of the indispensable steps in the natural language processing, which can transform a word into its root word. We carried out lemmatization for decisions using NLTK, an open source Python library for Natural Language Processing.

\textbf{3) Removing Stop Words:} Stop words refer to those common but convey little meaning words, such as ``and'' and ``that''. Removing stop words can help to reduce the size of vocabulary of our dataset and reduce the dimension of feature vectors of the decisions. In this work, we combined English stop words list provided by NLTK and several self-defined stop words, such as ``\textit{IMHO}'' that means ``\textit{in my humble opinion}'', to remove stop words in decisions.

\subsection{Feature Selection} \label{experiment fs}
In the feature selection step, we excluded terms that appear only once since they may contain less valid information for classification. Additionally, different percentages of selected features may also impact the performance of the automatic approach, thus we considered the percentage of the selected features as part of configurations and intended to seek the best configuration for classifying decisions from the Hibernate developer mailing list. Investigating the impact of all percentages (i.e., 1\% to 100\%) of the selected features for each combination of feature extraction techniques and ML classifiers is rather time-consuming. To simplify our experiment, the percentage of selected features was set from 10\% to 100\% (with a step of 10\%). Note that feature selection will make no difference when all the (100\%) features are selected.

\subsection{Feature Extraction and Classifier Training}\label{Feature Extraction and Classifier Training}
We conducted a vectorization process for decisions in our dataset and transformed them into vectors that could represent the features of decisions through feature extraction techniques. As elaborated in Section \ref{Feature Extraction Techniques}, BoW, TF-IDF, and Word2Vec were exploited to extract textual features. Then these vectors were used to train and evaluate the four base classifiers and five ensemble classifiers. Multinomial NB was chosen as the version of NB since it is better than other NB algorithms \cite{mccallum1998comparison}. Also, we normalized the feature vectors obtained by Word2Vec in the range of [0,1] since Multinomial NB fails when some of features are negative. The main hyperparameters of the four base classifiers are listed in Table \ref{Main Settings of Four Base Classifiers}. We used scikit-learn and gensim toolbox to implement the aforementioned feature extraction techniques and classifiers because they provide solid implementations of a bunch of state-of-the-art models.


\begin{table}[h]
\caption{Main Hyperparameters of Four Base Classifiers}
\label{Main Settings of Four Base Classifiers}
\begin{tabular}{@{}llll@{}}
\toprule
 & \textbf{Base classifiers} & \textbf{Main hyperparameters}              &  \\ \midrule
 & NB              & $alpha$: 1   &  \\
 & LR              & $C$: 1.0; $max\_iter$: 100; $solver$: lbfgs  &  \\
 & SVM             & $kernel$: linear; $C$: 1.0;  &  \\
 & RF              & $n\_estimators$: 100; $criterion$: gini      &  \\ \bottomrule
\end{tabular}
\end{table}

\subsection{Evaluation}\label{Evaluation}
Precision, recall, and F1-score (see Formula (\ref{precision-formula}), (\ref{recall-formula}) and (\ref{f1-formula})) were adopted as standard metrics to evaluate the performance of the aforementioned classifiers. Furthermore, we introduced weighted average precision, recall, and F1-score (see Formula (\ref{weighted-formula})), which are commonly used to measure the performance of classification results in imbalance datasets. To reduce overfitting in the classification of decisions, a \textbf{stratified 10-fold cross validation} was applied on the dataset, which is commonly used in machine learning as it produces less biased accuracy estimations for datasets with small sample sizes \cite{refaeilzadeh2009cross}. In addition, we repeated the 10-fold cross validation ten times and calculated the average results.

\begin{equation}
\label{precision-formula}
	Precision_{i}=\frac{TP_{i}}{TP_{i}+FP_{i}} 
\end{equation}

\begin{equation}
\label{recall-formula}
	Recall_{i}=\frac{TP_{i}}{TP_{i}+FN_{i}}
\end{equation}

\begin{equation}
\label{f1-formula}
	F1_{i}-score=\frac{2 \times Precision_{i} \times Recall_{i}}{Precision_{i}+Recall_{i}}
\end{equation}

\begin{equation}
\label{weighted-formula}
	\begin{aligned}
		Weighted \ (Precision \backslash Recall \backslash F1) = \\ \frac{\sum\limits_{i \in type}^{} (Precision_{i} \backslash Recall_{i} \backslash F1_{i}) \times Number_{i}}{\sum\limits_{i \in type}^{} Number_{i}}
	\end{aligned}
\end{equation}

In Formula (\ref{precision-formula}) and (\ref{recall-formula}), \bm{$TP_{i}$} denotes the number of decisions that are classified as type \bm{$i$} and actually belong to type \bm{$i$}; \bm{$FP_{i}$} denotes the number of decisions that are classified as type \bm{$i$} but actually belong to another type \bm{$j$} ($i$ $\neq$ $j$); \bm{$FN_{i}$} denotes the number of decisions that are of type \bm{$i$} but classified as another type \bm{$j$} ($i$ $\neq$ $j$). In Formula (\ref{weighted-formula}), \bm{$Number_i$} denotes the number of decisions labelled as decision type \bm{$i$} in the testing set.

\section{Results and Analysis}\label{Results and Analysis}
As described in Section \ref{Research Design}, we performed our experiment with 270 configurations (i.e, \textbf{10} different values of $n$ in features selection $\times$ \textbf{3} feature extraction techniques $\times$ \textbf{9} ML classifiers). For each configuration, we calculated the metrics (i.e., weighted precision, recall, and F1-score). In this section, we report the experiment results and answer the RQs. 

\textbf{To answer RQ1:} To seek whether feature selection improves the classification results, we compared the classification results for all combinations of feature extraction techniques and classifiers with and without feature selection as shown in Table \ref{fsc}, in which FS denotes Feature Selection. Especially, we listed the best F1-score and the corresponding value of $n$ for each combination when the selected features vary from 10\% to 90\% (See Section \ref{experiment fs}). It can be found that the F1-scores of all combinations increase when applying feature selection. The lowest improvement (1.9\%) is achieved by BoW + RF, whereas Word2Vec + RF achieves the highest improvement (40.6\%). The improvements also differ from different feature extraction techniques. For BoW, the F1-scores are slightly increased with an average improvement of 5.3\%. For TF-IDF, the results are better compared with BoW. On average, the feature selection improves F1-score by 6.6\%. The F1-scores of Word2Vec are significantly improved compared with BoW and TF-IDF. The average improvement reaches 25.9\%. Another important observation is that for the same feature extraction technique, the percentages of selected features (i.e., the value of $n$) are close when most classifiers achieve the best result. For example, 50\% for BoW, and 20\%$\sim$40\% for TF-IDF. Moreover, nine classifiers with Word2Vec all achieve the best F1-score when 10\% features are selected since the F1-score will decrease when we increase the percentage of selected features. Besides the improvement on classification results, feature selection can provide extra performance since it can reduce the dimensions of data on which we need to train. \textbf{In summary, feature selection has an impact on the classification results and can decently improve the classification results when classifying decisions from the Hibernate developer mailing list.}

\begin{table}[h]
\caption{Comparison of F1-scores of All Combinations of Feature Extraction Techniques and Classifiers with and without Feature Selection}
\label{fsc}
\begin{tabular}{llll}
\hline
\textbf{Combinations} & \textbf{NoFS} & \textbf{FS ($n$)} & \textbf{Improve} \\ \hline
BoW + NB              & 0.653         & 0.709 (40\%)      & 8.6\%            \\
BoW + LR              & 0.678         & 0.713 (50\%)      & 5.2\%            \\
BoW + SVM             & 0.650         & 0.715 (50\%)      & 10.0\%           \\
BoW + RF              & 0.645         & 0.657 (10\%)      & 1.9\%            \\
BoW + SVE1            & 0.695         & 0.722 (50\%)      & 3.9\%            \\
BoW + SVE2            & 0.689         & \textbf{0.727} (50\%)      & 5.5\%            \\
BoW + SVE3            & 0.680         & 0.712 (50\%)      & 4.7\%            \\
BoW + SVE4            & 0.693         & 0.722 (50\%)      & 4.2\%            \\
BoW + SVE5            & 0.695         & 0.719 (50\%)      & 3.5\%            \\
Avg                   & 0.675         & 0.711             & 5.3\%            \\
                      &               &                   &                  \\
TF-IDF + NB           & 0.584         & 0.684 (20\%)      & 17.1\%           \\
TF-IDF + LR           & 0.640         & 0.703 (20\%)      & 9.8\%            \\
TF-IDF + SVM          & 0.684         & 0.716 (50\%)      & 4.7\%            \\
TF-IDF + RF           & 0.640         & 0.672 (40\%)      & 5.0\%            \\
TF-IDF + SVE1         & 0.690         & 0.720 (40\%)      & 4.3\%            \\
TF-IDF + SVE2         & 0.686         & 0.720 (30\%)      & 5.0\%            \\
TF-IDF + SVE3         & 0.693         & \textbf{0.724} (40\%)      & 4.5\%            \\
TF-IDF + SVE4         & 0.692         & 0.723 (40\%)      & 4.5\%            \\
TF-IDF + SVE5         & 0.656         & 0.700 (20\%)      & 6.7\%            \\
Avg                   & 0.663         & 0.707             & 6.6\%            \\
                      &               &                   &                  \\
Word2Vec + NB         & 0.377         & 0.500 (10\%)      & 32.6\%           \\
Word2Vec + LR         & 0.568         & \textbf{0.683} (10\%)      & 20.2\%           \\
Word2Vec + SVM        & 0.582         & \textbf{0.683} (10\%)      & 17.4\%           \\
Word2Vec + RF         & 0.461         & 0.648 (10\%)      & 40.6\%           \\
Word2Vec + SVE1       & 0.533         & 0.673 (10\%)      & 26.3\%           \\
Word2Vec + SVE2       & 0.555         & 0.663 (10\%)      & 19.5\%           \\
Word2Vec + SVE3       & 0.563         & 0.677 (10\%)      & 20.2\%           \\
Word2Vec + SVE4       & 0.515         & 0.667 (10\%)      & 29.5\%           \\
Word2Vec + SVE5       & 0.501         & 0.665 (10\%)      & 32.7\%           \\
Avg                   & 0.517         & 0.651             & 25.9\%           \\ \hline
\end{tabular}
\end{table}

\textbf{To answer RQ2:} To seek whether ensemble classifiers outperform the base classifiers, we analyzed the classification results of base classifiers and ensemble classifiers in Table \ref{fsc}. For BoW and TF-IDF, most ensemble classifiers perform better than base classifiers no matter whether feature selection is applied. Moreover, the best ensemble classifier outperforms the best base classifier (e.g., BoW + SVE2 + 50\% features achieves higher F1-score than BoW + SVM + 50\% features). However, not all ensemble classifiers are better than base classifiers, e.g., the F1-score of TF-IDF + SVM + 50\% features (0.716) is better compared with the F1-score of TF-IDF + SVE5 + 20\% (0.700). For Word2Vec, we can learn that the results are different. The classification results obtained by LR and SVM are better than the results of the five ensemble classifiers, whereas the classification results of NB and RF are worst. Since an ensemble classifier combines the predictions from base classifiers into the final prediction, performance of the ensemble classifier is thus influenced by the base classifiers. The base classifiers with good performance could benefit to construct a well-performed ensemble classifier while the base classifiers with bad performance would be harmful. For instance, NB and RF perform significantly worse than SVM and LR when using Word2Vec without feature selection, which results in the worse performance of five ensemble classifiers. The results of NB, LR, and SVM are relatively good when using BoW with feature selection, which results in that the ensemble classifier that combines these three base classifiers achieve a better classification result. \textbf{In summary, the ensemble classifiers may outperform base classifiers, but it depends on how these ensemble classifiers are constructed when classifying decisions from the Hibernate developer mailing list.} It indicates that the construction of ensemble classifiers should also be considered when using ensemble learning in automatic classification tasks. In order to make the ensemble learning more valuable in practice, it is necessary to seek suitable base classifiers, which is also our next step.   

\textbf{To answer RQ3:} To determine the best configuration that achieves the best performance to classify decisions from the Hibernate developer mailing list, we compared and ranked all configurations based on their weighted F1-scores. We listed the top 5 configurations in Table \ref{rbc}. It can be found that the F1-scores of the top 5 configurations are all higher than 0.72. We can also learn that the highest precision (i.e., 0.754) and recall (i.e., 0.739) are achieved by TF-IDF + SVE4 + 40\% features and BoW + SVE2 + 50\% features, respectively. Overall, the BoW + SVE2 + 50\% features obtained the best classification result with a weighted F1-score of 0.727. In addition, the top 5 configurations have some characteristics in common: (1) they all applied feature selection; (2) they all employed ensemble classifier; and (3) they do not utilize Word2Vec to extract features of decisions. 

The results in Table \ref{rbc} indicate that (1) feature selection has an advantage when classifying decisions from the Hibernate developer mailing list, which is consistent with our conclusion of RQ1; (2) ensemble classifiers could achieve better classification results than base classifiers, which confirms our conclusion of RQ2; (3) BoW and TF-IDF are more suitable feature extraction techniques compared with Word2Vec in our cases. This finding is in line with the result in \cite{liautomatic}, which also focused on decisions from the Hibernate developer mailing list. One possible reasons could be that the size of our dataset is limited. Thus, limited information is provided for Word2Vec to learn enough semantic knowledge, which may lead to the poor performance of Word2Vec. \textbf{In summary, BoW + SVE2 + 50\% features outperforms other configurations. It can automatically classify decisions from the Hibernate developer mailing list with a weighted precision of 0.750, a weighted recall of 0.739, and a weighted F1-score of 0.727.}

\begin{table}[h]
\caption{Results of the Best 5 Configurations for Classifying Decisions}
\label{rbc}
\begin{tabular}{llll}
\hline
\textbf{Configurations}       & \textbf{Precsion} & \textbf{Recall} & \textbf{F1} \\ \hline
BoW + SVE2 + 50\% features    & 0.750              & \textbf{0.739}  & \textbf{0.727}    \\
TF-IDF + SVE3 + 40\% features & 0.753              & 0.737           & 0.724             \\
TF-IDF + SVE4 + 40\% features & \textbf{0.754}     & 0.737           & 0.723             \\
BoW + SVE1 + 50\% features    & 0.747              & 0.734           & 0.722             \\
BoW + SVE4 + 50\% features    & 0.745              & 0.734           & 0.722             \\ \hline
\end{tabular}
\end{table}

\section{Discussion}\label{Discussion}
\subsection{Feature Terms after Feature Selection}
In our experiment, we use feature selection to select terms that may contain valid type information for classification. We present the top 20 feature terms in Table \ref{feature words} after feature selection for each decision type. We focus on seeking which terms may provide expected discrimination between types. However, since the number of testing decisions in our dataset is limited (i.e., 23 of 844) and we excluded terms that appear only once in feature selection (see Section \ref{experiment fs}), there are less than 20 terms in Testing Decision. Note that the CHI$^2$ value indicates the amount of information carried by a term for predicting the type of a decision, but a term with a high CHI$^2$ value of one decision type does not necessarily mean that the decision with this term belongs to this type. 

To better understand these feature terms, we also manually checked each term in Table \ref{feature words} and highlighted terms that are highly relevant to each decision type by the first and second authors. Any conflicts were discussed and resolved among all the authors. From the results, we can find that feature selection can pick out the terms that are highly related to the definition of each decision type (see Section \ref{Types of Decisions}). For \textbf{Design Decision}, to best of our knowledge, terms like ``move'', ``refactor'', and ``design'' may be highly correlated to design decisions (e.g., ``\textit{He has started refactoring SessionImpl, as the first step of redesigning Hibernate to be more event oriented, as we all discussed a while ago}''). However, there are also feature terms that are not directly relevant to the design of software systems, such as ``session'', ``exception'', ``transaction'', and ``bean''. Since Hibernate, an Object/Relation Mapping (ORM) library for java projects, is used as the source of our dataset, these terms are specific concepts in the Hibernate and would be involved when Hibernate designers make decisions. For example, session is used to get a physical connection with a database, and it plays a major role between the database and Hibernate. Therefore, ``session'' is emphasized in the design decisions, such as ``\textit{Keep a single session in a ThreadLocal, or the DAOFactory, to be shared by all DAO instances in that transaction}''.

``support'', ``feature'', ``request'', and so on are feature terms of \textbf{Requirement Decision}. Decisions with these terms are usually related to developers improving the requirements on functionalities, such as ``\textit{There should also be a feature enquiry function to allow the API programmer to ask the underlying Hibernate implementation to verify if a replication on identity table will work}''. Besides, ``performance'', as one of the common non-functional requirements, is also highly related to Requirement Decision. Hibernate developers would make decisions to improve the quality of Hibernate products, as well as ``\textit{a performance gain could be achieved (on some databases) by selectively updating columns}''. 

For \textbf{Management Decision}, ``documentation'', ``javadoc'', and so on are the terms that indicate the maintenance and improvements in documentations of Hibernate, such as ``\textit{The new internationalized documentation will be in 2.1.3, so we have to update the Wiki documentation page before release}'', which meets the definition of Management Decision. In the meanwhile, ``release'' is the term with the highest CHI$^2$ value for Management Decision. The correlative decisions are made when a new version of Hibernate is completed and ready to release, such as ``\textit{We will release Hibernate 3.0alpha early next week}''. 

In addition, most feature terms of \textbf{Construction Decision} are related to the source code and configuration files of the Hibernate, such as ``override'', ``annotation'', and ``buildxml'', since construction decisions focus on the implementation details of the software systems. Hibernate developers would discuss the code issues with each other in the mailing list, such as ``\textit{I have also overriden the hashCode() and equals() methods on StuffUnitFactor so that they are equal when stuff \&\& unit are equal}''. 

As for \textbf{Testing Decision}, ``test'' is the most frequent term and it is always related to the testing activities during the development and maintenance of software, such as ``unit test'' and ``test cases'' (e.g., ``\textit{I will add a test case for the new Transaction API to the FooBarTests as soon as I get a chance}'').

We also find that different decision types may share some common terms, such as ``add'' and ``use''. Most of these terms are verbs and are frequently used in the communication between stakeholders. For example, Hibernate developers would use linguistic patterns like ``add support'' and ``add ability'' to express the improvement on requirements.

\begin{table*}[]
\caption{Top 20 Feature Terms after Feature Selection for Each Decision Type}
\label{feature words}
\begin{tabular}{@{}lllllll@{}}
\toprule
 & \textbf{Design Decision} & \textbf{Requirement Decision} & \textbf{Management Decision} & \textbf{Construction Decision} & \textbf{Testing Decision} &  \\ \midrule
 & add                  & \textbf{support}     & \textbf{release}       & \textbf{override}   & \textbf{test}    &  \\
 & \textbf{session}     & add                  & \textbf{documentation} & \textbf{line}       & suite            &  \\
 & use                  & \textbf{feature}     & \textbf{page}          & \textbf{parameter}  & \textbf{unit}    &  \\
 & support              & \textbf{implement}   & \textbf{wiki}                   & \textbf{code}       & run              &  \\
 & join                 & \textbf{performance} & \textbf{final}                  & equal               & \textbf{testing} &  \\
 & \textbf{move}                 & write                & \textbf{readme}                 & \textbf{classpath}  & \textbf{case}    &  \\
 & application          & \textbf{hql}         & \textbf{document}      & \textbf{annotation} & order            &  \\
 & \textbf{exception}   & code                 & \textbf{doc}           & \textbf{lib}        & \textbf{fix}     &  \\
 & place                & criterion            & \textbf{vbranch}       & hibernateproperties & \textbf{functionality}    &  \\
 & way                  & full                 & hibernateext           & false               & get              &  \\
 & \textbf{transaction} & \textbf{request}     & reference              & unsavedvalue        & cvs              &  \\
 & \textbf{remove}               & use                  & week                   & embrace             & need             &  \\
 & \textbf{refactor}    & generate             & website       & style               & cache            &  \\
 & \textbf{bean}                 & \textbf{task}        & section                & figure              & new              &  \\
 & \textbf{design}      & eclipse              & note                   & onetomany           & file             &  \\
 & test                 & custom               & \textbf{branch}        & message             & use              &  \\
 & put                  & syntax               & \textbf{guide}         & finally             & make             &  \\
 & \textbf{solution}    & test                 & distribution           & cut                 & add              &  \\
 & return               & select               & jar                    & extra               & change           &  \\
 & directly             & provide     & \textbf{javadoc}       & \textbf{bulidxml}            &                  &  \\ \bottomrule
\end{tabular}
\end{table*}

\subsection{Implications for Practitioners}
Our approach can benefit various types of stakeholders in software development (e.g., project managers, requirements engineers, designers, and developers), because it could effectively classify decisions into specific types that are useful for them. Since there are approaches focusing on the automatic identification of decisions (e.g., \cite{liautomatic}) but lacking the multiclass classification methods of decisions (not only design decisions), our approach could help project managers better document decisions with minimal human effort. Then other stakeholders could review, analyze, improve, and reuse the decisions that they are concerned with. For instance, designers could gain knowledge from past design decisions, such as ideas and insights toward the design issues, which could help them make better decisions in a similar design context. Requirement engineers could be aware of what decisions have been made toward the functional and non-functional requirements, and then determine new requirements to improve the software system. Developers can be aware of the impact of software changes caused by related decisions, which helps the developers to avoid making poor decisions that lead to bug proneness and increased maintenance effort \cite{paixao2017developers}. Moreover, our approach could inform the newcomers of the decisions that are highly related to their work so that they can have a better understanding of the software system and get involved in the work more efficiently. Also, our approach is also scalable in practice. Project managers could re-select feature terms and retrain our approach while more decisions are classified.


\subsection{Implications for Researchers}
Decisions, as a type of inherent artifact in software development and maintenance, get widespread attention from researchers in recent years (e.g., \cite{mendes2019relationship}). Many researches have focused on analyzing decisions, especially design decisions, and their making in software development. In this work, we adopted the taxonomy from our previous research \cite{li2019decisions} that classifies decisions in software development into five types, which can basically cover the decisions in the Hibernate developer mailing list. We further adapted the description of the five types of decisions and provided the decision dataset online \cite{replication-package}. This dataset can be extended with the decisions from other projects and sources, acting as a foundation for other researchers to work on automatic classification of decisions. We also propose an automatic approach (i.e., the BoW + 50\% features selected by CHI$^2$ algorithm with an ensemble classifier that combines NB, LR, and SVM) to classify decisions. Although our approach is evaluated on one OSS project (i.e., Hibernate), we believe that our approach could provide the basis for designing and developing tools to automatically classify decisions, with the knowledge and usage of feature selection and ensemble learning. In addition, considering that the research in the area of automatic decision classification is still in its infancy, we advocate that researchers can pay more attention to the design of automatic and interactive tools for classifying decisions in software development.

\section{Threats to Validity}\label{Threats to Validity}
In this section, we discuss several threats to the validity of this work according to the guidelines proposed by Runeson \textit{et al.} \cite{runeson2009guidelines}, and how these threats were partially mitigated in our study. Internal validity is not considered, since we do not address causal relationships between variables and results.

\textbf{Construct Validity} reflects on the extent of consistency between the operational measures of the study and the RQs \cite{runeson2009guidelines}. A potential threat in this study involves whether the decisions used for the experiment were classified correctly by the researchers. To mitigate this threat, we reviewed and labelled the decisions by two researchers. Any disagreements were discussed and resolved with another researcher. Moreover, we conducted a pilot labeling with 300 decisions to mitigate any personal bias and achieve a consensus understanding on manually labelling. Another potential threat in this study is whether the dataset for the experiment is sufficient enough to obtain reasonable conclusions since our dataset only contains 844 decisions and is also quite imbalanced. To mitigate this threat and get reasonable results, we repeated stratified 10-fold cross validation 10 times with three weighted metrics (i.e., weighted precision,  weighted recall, and weighted F1-score) to evaluate our experiment results. In addition, Manning \textit{et al.} argued that the classification performance will be improved when increasing the sample size of the dataset \cite{manning1999foundations}. Therefore, we conjecture that we can obtain more accurate conclusions by increasing the sample size of the dataset and making it balanced to mitigate the threat of the insufficient dataset, which is also our next research goal. 

\textbf{External Validity} refers to the degree to which our study results and findings can be generalized in other cases (e.g., the developer mailing lists of other OSS projects). We collected decisions from the developer mailing list of Hibernate since it is one of the most popular OSS projects and is widely used in industry as an ORM library for Java projects. However, we acknowledge that our study may not be generalizable to other OSS projects. For instance, the feature terms acquired for Hibernate may not be applicable to other projects. Different projects can lead to different feature terms since these terms may be related to a variety of aspects, such as programming language and domain of the project. But we believe that our work can promote researches on classifying decisions into types that are beneficial for various types of stakeholders in software development. In the next step, we plan a study through extending our dataset of decisions from more OSS projects with various characteristics (e.g., other programming languages and domains), which will alleviate this threat.

\textbf{Reliability} refers to whether the experiment yields the same results when it is replicated by other researchers. To improve the reliability, we made a research protocol and present all the details of the experiment in Section \ref{Experiment}. Furthermore, the dataset and the source code of our experiments are made publicly available online in order to facilitate other researchers replicating our experiment easily \cite{replication-package}. We believe that this measure can partially alleviate this threat.

\section{Conclusions and Future Work}\label{Conclusions and Future Work}
Decisions are a typical type of software artifacts and they have a crucial impact on development and maintenance of software systems. Nevertheless, decisions are not well documented due to considerable human resources, time, and budget. To address this issue, we conduct an experiment to automatically classify decisions from the Hibernate developer mailing list into five decision types (i.e., Design, Requirement, Management, Construction, and Testing Decision), which were got in our previous work using grounded theory \cite{li2019decisions}.

In this work, we first collected our dataset with 844 decisions of five decision types. To obtain the best approach for classifying decisions, we performed our experiment with 270 configurations. In feature selection step, we varied the percentages of the selected features from 10\% to 100\% (with a step of 10\%). Three well-performed feature extraction techniques (i.e., BoW, TF-IDF, and Word2Vec) were used to extract features of decisions. As for classifiers, we utilized an ensemble learning approach and constructed five ensemble classifiers using four base ML classifiers (i.e., NB, LR, SVM, and RF). We also compared the performance between base classifiers and ensemble classifiers. All the configurations were evaluated through weighted precision, recall, and F1-score. The results show that (1) feature selection can decently improve the classification results; (2) ensemble classifiers can outperform base classifiers provided that ensemble classifiers are well constructed; (3) BoW + 50\% features selected by feature selection with an ensemble classifier that combines NB, LR, and SVM achieves the best classification result (with a weighted precision of 0.750, a weighted recall of 0.739, and a weighted F1-score of 0.727) among all the configurations. Our approach can benefit various types of stakeholders in software development (e.g., project managers, requirements engineers, designers, and developers) by classifying decisions into specific types that are relevant to their interests.

Given the importance of decisions in software development, we plan to: (1) collect decisions from other OSS projects to get a larger and balanced dataset, and then validate our approach on these projects; (2) evaluate other state-of-art techniques (e.g., FastText and Text-CNN) to improve the performance of our approach on classifying decisions; (3) use automatic approaches to mine diverse information from decisions (e.g., rationale of decisions) to enrich the documentation of decisions; and (4) use automatic approaches to classify other types of software artifacts related to decisions, such as assumptions, to promote the understanding and (re)use of decisions in software development.

\begin{acks}
This work has been partially supported by the National Key R\&D Program of China with Grant No. 2018YFB1402800.
\end{acks}

\balance
\bibliographystyle{ACM-Reference-Format}
\bibliography{reference}

\end{document}